\documentclass[twocolumn,showpacs,preprintnumbers,amsmath,amssymb,prl,superscriptaddress]{revtex4}

\usepackage{graphicx}% Include figure files
\usepackage{dcolumn}% Align table columns on decimal point
\usepackage{bm}% bold math

\begin{document}

%\preprint{APS/123-QED}

\title{Intrinsic susceptibility and bond defects in the novel 2D frustrated antiferromagnet Ba$_{2}$Sn$_{2}$ZnCr$_{7p}$Ga$_{10-7p}$O$_{22}$}

\author{D.~Bono}
\author{P.~Mendels}%
% \email{mendels@lps.u-psud.fr}
\affiliation{%
Laboratoire de Physique des Solides, UMR 8502, Universit\'e Paris-Sud, 91405 Orsay, France
}%

\author{G.~Collin}
\affiliation{
Laboratoire L\'eon Brillouin, CE Saclay, CEA-CNRS, 91191 Gif-sur-Yvette, France
}%

\author{N.~Blanchard}%
\affiliation{%
Laboratoire de Physique des Solides, UMR 8502, Universit\'e Paris-Sud, 91405 Orsay, France
}%

\date{\today}

\begin{abstract}
We present microscopic and macroscopic magnetic properties of the highly frustrated antiferromagnet Ba$_{2}$Sn$_{2}$ZnCr$_{7p}$Ga$_{10-7p}$O$_{22}$, respectively probed with NMR and SQUID experiments. The $T$-variation of the intrinsic susceptibility of the Cr$^{3+}$ frustrated kagom\'{e} bilayer, $\chi_{kag}$, displays a maximum around 45~K. The dilution of the magnetic lattice has been studied in detail for $0.29 \leq p \leq0.97$. Novel dilution independent defects, likely related with magnetic bond disorder, are evidenced and
discussed. We compare our results to SrCr$_{9p}$Ga$_{12-9p}$O$_{19}$. Both bond defects and spin vacancies do not affect\ the average susceptibility of the kagom\'{e} bilayers. 
\end{abstract}

%\keywords{frustrated magnet, kagom\'e}%Use showkeys class option if keyword
\maketitle

Geometric frustration in magnetism has proven to yield various original ground states in the past decade, from RVB ``spin liquids'', to exotic freezings as spin ice. Most of these systems have in common a simple Heisenberg Hamiltonian with near neighbor (\emph{nn}) antiferromagnetic interactions on a corner sharing lattice, \emph{e.g.} the kagom\'e lattice in two dimensions~\cite{Ramirez01}. In the latter case and $S=\frac{1}{2}$, the ground state is expected to be the long sought resonating valence bond (RVB) state~\cite{Mambrini00}. This opens not only the way to novel states in low dimensional quantum magnetism~\cite{Diep03} but the RVB state has been also advocated in many fields of condensed matter such as high $T_{c}$ cuprates~\cite{Anderson87} or recently discovered cobaltites, where frustration is inherent to the triangular Co network~\cite{Baskaran03}. Geometric frustration is also a rapidly developing field since in the classical Heisenberg case, small perturbations such as disorder, anisotropy or weak additional interactions, lift the frustration-induced macroscopic degeneracy of the ground state~\cite{Chubukov92}. 

So far, there is no ideal experimental candidate for a spin liquid behavior, but a few, like the newly discovered Ba$_{2}$Sn$_{2}$ZnCr$_{7p}$Ga$_{10-7p}$O$_{22}$ compound (BSZCGO$(p)$)~\cite{Hagemann01}, retain the essence of the expected features, \emph{e.g.} a short correlation length~\cite{BonnetHFM} which prevents the occurence of any long range order in a broad $T$-range below the Curie-Weiss (CW) temperature $\theta_{CW}$, a large residual entropy~\cite{Hagemann01}, and persistent fluctuations at $T\rightarrow 0$~\cite{BonoPrep}. The corner sharing magnetic lattice of BSZCGO$(p)$ is made of well decoupled Cr$^{3+}$ ($S=\frac{3}{2}$) 2D kagom\'{e} bilayers \emph{only}. Interestingly, the \emph{nn} direct couplings between Cr$^{3+}$ ions make the Hamiltonian quite close to the \emph{ideal Heisenberg} case~\cite{Limot02}. A spin-glass like transition occurs at $T_{g}\approx1.5$~K~\cite{Hagemann01}, well below $\theta_{CW}\approx350$~K, which yields the highest \emph{frustration ratio} $f=\theta_{CW}/T_{g}\sim230$ reported so far in a compound where frustration is driven by corner sharing equilateral triangles~\cite{Ramirez01}.

In comparison, the long-studied parent Cr$^{3+}$ compound SrCr$_{9p}$Ga$_{12-9p}$O$_{19}$ (SCGO$(p)$)~\cite{Obradors88} ($f\sim150$) is not made of frustrated bonds only, since $\frac{2}{9}$ of the Cr$^{3+}$ form pairs which separate the bilayers and couple into singlets below $216$~K~\cite{Lee96}. Ga$^{3+}$/Cr$^{3+}$ substitutions of the order of a few percent ($\equiv 1-p$) represent the major drawback in SCGO$(p)$ since they introduce paramagnetic centers in between the bilayers by breaking these Cr$^{3+}$ pairs. This not well-controlled parameter, together with the shorter inter-bilayer distance, 6.4~\AA~\cite{Lee96} instead of 9.4~\AA \ in BSZCGO$(p)$~\cite{Hagemann01}, is likely to make the physics of SCGO$(p)$ less 2D and complicates the search for the intrinsic kagom\'e physics.

We present the first NMR study of BSZCGO$(p)$ and show that the intrinsic susceptibility ($\chi_{kag}$) displays a maximum at a temperature comparable to the \emph{nn} Heisenberg coupling $J$. The similarity of the $T$-variation of $\chi_{kag}$ with the case of SCGO$(p)$ stresses the universal character of this maximum in $\chi_{kag}$ for kagom\'e bilayers. As a \emph{local} probe, Ga NMR, contrary to SQUID, offers a unique way to disentangle $\chi_{kag}$ from the contribution of ``magnetic defects'' to macroscopic susceptibility~\cite{Mendels00,Limot02}. This gives evidence for two types of defects, for $0.29 \leq p \leq0.97$, $(i)$~the spin vacancies due to the Ga$^{3+}$/Cr$^{3+}$ substitution which was shown to generate a staggered response of the surrounding magnetic background~\cite{Limot02} and $(ii)$~ novel dilution independent defects, discussed here.

All the samples were synthesized by a solid state reaction of BaCO$_{3}$, Cr$_{2}$O$_{3}$, Ga$_{2}$O$_{3}$, ZnO and SnO$_{2}$ in air at 1\,400$^{\circ}$C. They were characterized by X-ray diffraction, macroscopic susceptibility ($\chi_{macro}$) measurements and $^{71}$Ga NMR. The synthesis of pure samples with $p>0.97$ failed. Indeed, in this case, X-ray diffraction shows a Cr$_{2}$O$_{3}$ parasitical phase and an extra kink of $\chi_{macro}$ is also observed around 6~K. This sets the maximum of the Cr$^{3+}$ lattice occupancy to a value of $p_{max}=0.97$. Below $p_{max}$, Ga substitutes randomly on the Cr(1a, 6i) site. 

\begin{figure}[tb]
\center
\includegraphics[width=0.9\linewidth]{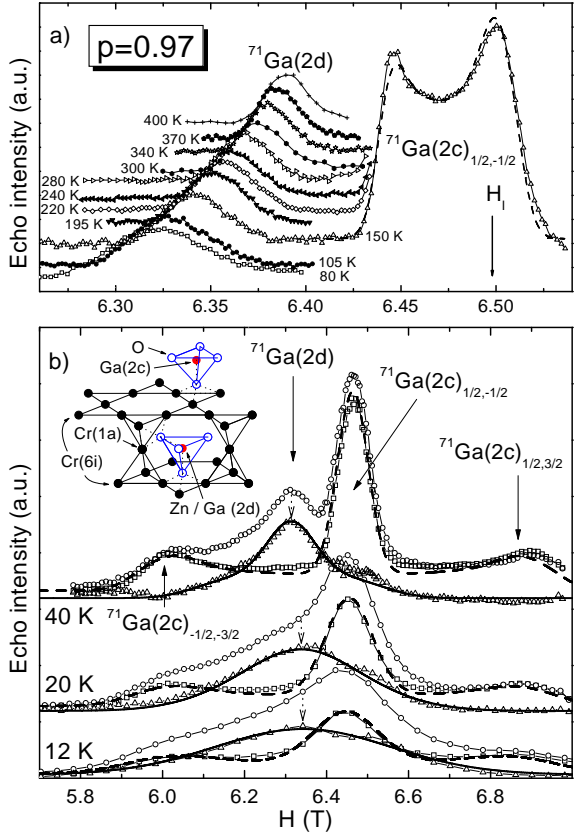}
\caption{$^{71}$Ga spectra. a)~$T\geq80$~K. The $^{71}$Ga(2c) first order quadrupolar contribution appears as a flat background in this field range. b)~$T\leq40$~K, lines are broadened. $\circ$ and $\square$ are the short $\tau$ and the rescaled long $\tau$ spectra respectively. $\triangle$ is for the Ga(2d) contribution, given by their subtraction. The dotted arrows point the center of the Ga(2d) line. Continuous (dashed) lines are gaussian broadened $^{71}$Ga(2d) ($^{71}$Ga(2c)) powder pattern quadrupolar simulation, with $\nu_{Q}=3.5$~MHz (12~MHz) and $\eta=0.6$ ($\eta=0.04$). Inset~: Ga$^{3+}$ sites. }%
\label{spectres}%
\end{figure}

The two Ga$^{3+}$ site environments ($p=1$), are displayed in the inset of Fig.~\ref{spectres}(b). The Ga(2d) site is coupled to $3p$ Cr(1a) and $9p$ Cr(6i) through neighboring oxygens. Its local charge environment is approximately cubic, whereas the Ga(2c) site belongs to a tetrahedron elongated along the $c$-axis and couples to $3p$ Cr(6i).

We performed $^{71}$Ga sweep field NMR experiments at constant frequency $\nu_{l}=84.365$~MHz, using a $\frac{\pi}{2}$-$\tau$-$\pi$ spin echo sequence.
Two sites can be clearly identified in Fig.~\ref{spectres} with the expected ratio of integrated intensities close to 2:1. At 150~K, the double peak around the reference field $H_{l}\approx6.498$~T in Fig~\ref{spectres}(a) is characteristic of a 0.53\% slightly shifted $\frac{1}{2}\leftrightarrow-\frac{1}{2}$ nuclear transition, with second order quadrupolar powder pattern. The corresponding first order satellite singularities are clearly seen in the less expanded spectra of Fig.~\ref{spectres}(b). From these singularities, we extract a quadrupolar frequency $^{71}\nu_{Q}\approx
12$~MHz and an asymmetry parameter $\eta=0.04(2)$~\cite{Keren98}. The other line is more shifted $\sim2$\% (Fig.~\ref{spectres}(a)). Its shape can be explained by a broadened $S=\frac{3}{2}$ quadrupolar pattern, with $^{71}\nu_{Q}= 3.5\pm1$~MHz (Fig.~\ref{spectres}(b))~\footnote{Using a point charge computation, we could estimate a 30\% distribution of $\nu_{Q}$ and of $\eta\sim0.6$, due to the Zn$^{2+}$/Ga$^{3+}$ random occupation of the 2d site. It prevents any observation of quadrupolar satellites at high-$T$.}. The intensity ratio, the very different quadrupolar frequency associated with different charge environment and the analogy with SCGO allow us to identify unambiguously the Ga(2c) and Ga(2d) sites. 
Though, contrary to SCGO$(p)$, we could not identify any line typical of substituted Ga.

At low-$T$, the two lines are found to overlap due to strong broadenings. In order to deconvolute the two contributions, we took advantage of the different transverse relaxation times $T_{2}$ of the two lines
(Fig.~\ref{spectres}b). The first step was to use a large $\tau$ spectrum ($\tau=200~\mu$s) to isolate the Ga(2c) contribution ($T^{71}_{2,2c}\approx200~\mu$s). Its linewidth $\Delta H$ is extracted using a gaussian convolution of the high-$T$ quadrupolar pattern. In the second step, we used the short $\tau$ spectrum ($\tau\approx10~\mu$s) where the Ga(2d) contribution is evident, and subtracted the $T^{71}_{2,2d}$ normalized long $\tau$ spectrum. The width and the shift of the Ga(2d) line can then be extracted from a fit using a $^{71}\nu_{Q}=3.5$~MHz quadrupolar pattern convoluted by a gaussian line shape, not affected by the 30\% uncertainty on $^{71}\nu_{Q}$. As a matter of consistency, the ratio $I^{71}_{2c}/I^{71}_{2d}=2\pm0.8$ stays close to $2/1$ down to 10~K. No significant loss of intensity was observed down to this temperature.

\begin{figure}[tb]
\center
\includegraphics[width=1\linewidth]{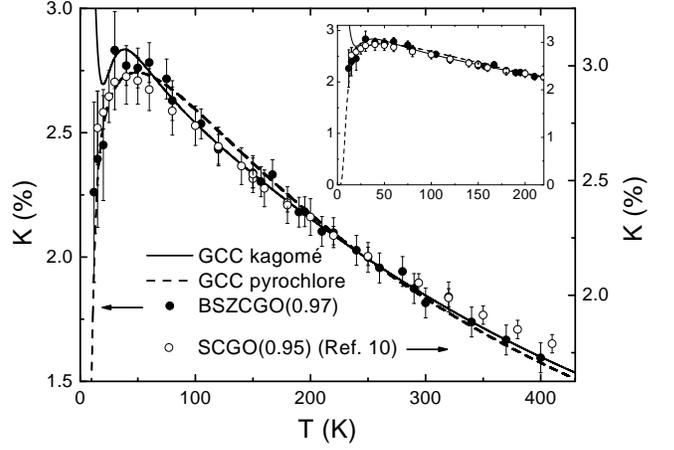}
\caption{Shift of the $^{71}$Ga(2d) (Ga(4f)) line for BSZCGO (SCGO). Inset~: full scale down to 0. Lines are fits (see text). }%
\label{shift}%
\end{figure}

The kagom\'e bilayer susceptibility $\chi_{kag}$ is revealed in the shift $K_{2d}^{71}$ of the Ga(2d) line~\cite{Mendels00}, presented in Fig.~\ref{shift}. It increases when $T$ decreases for $T\geq80$~K, reaches a maximum around $T_{max}\approx45$~K and decreases below. A phenomenological CW law, $K_{2d}^{71}=C_{NMR}/(T+\theta_{NMR})$, is an accurate fit for $T>80$~K with $\theta _{NMR}=380\pm10$~K. Although very common, this is a rather crude approximation in a $T$-range of the order of $\theta_{CW}$.
In the absence of any prediction for $S=\frac{3}{2}$, we used the high-$T$ series expansion of $\chi_{kag}$, derived for a $S=\frac{1}{2}$ kagom\'e lattice, to correct $\theta_{NMR}$ by a factor 1.5~\cite{Harris92}.
From $\theta_{CW}=zS(S+1)J/3$ and $z=5.14$ the average number of \emph{nn} for the Cr$^{3+}$ of the bilayer, we extract $J\approx40$~K. 

In SCGO$(0.95)$~\cite{Limot02}, a value $\theta_{NMR}=440\pm5$~K~\footnote{The difference in $\theta_{NMR}$ with \cite{Limot02} comes from the chemical shift which was neglected
there. We estimated it here to 0.15(3)\% comparing $K_{2d}^{71}$ and $K_{2c}^{71}$ at high-$T$.}, was found and $C_{NMR}$ is 20\% larger. The small variation in the CW behaviors can be explained by a slightly larger Cr$^{3+}$-Cr$^{3+}$ coupling, due to a  0.01~\AA\ mean shorter bond, and $\Delta J /\Delta d\sim450$~K/\AA~\cite{Limot02}.

\begin{figure}[tb]
\center
\includegraphics[width=0.9\linewidth]{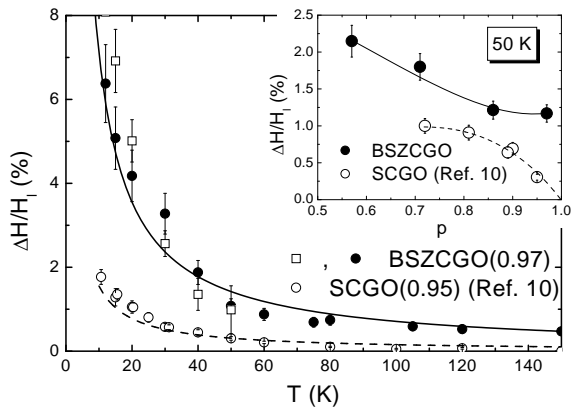}
\caption{Magnetic contribution to the NMR linewidth $\Delta H/H_{l}(^{71}$Ga(2d)) ($\bullet$). Data for the $^{71}$Ga(2c) ($\square$) are rescaled by a factor 6, corresponding to the ratio of the coupling constants, consistent with the high-$T$ shifts. The lines are $\propto1/T$ fits. Inset~: $p$-dependence of $\Delta H^{71}_{2d}$ ($\Delta H^{71}_{4f}$) at 50~K for BSZCGO$(p)$ (SCGO$(p)$). The lines are guides to the eye. }%
\label{larg}%
\end{figure}

The existence of a gap $\Delta$ in $\chi_{kag}$ was already an important issue for SCGO$(0.89)$ \cite{Mendels00}, since it would be a unique signature of the existence of a singlet ground state. For $S=\frac{1}{2}$, its value is predicted to be $\Delta\sim J/20$~\cite{Waldtmann98}, which should be little influenced by the spin value~\footnote{C.~Lhuillier, \emph{private communication}.}. Its definite observation would then require $T<\Delta$ in order to observe the exponential decrease of $\chi_{kag}$. Since our data is masured above $J/4\sim10$~K, we only tentatively fitted $K_{2d}^{71}$ with the phenomenological function $\propto\exp(-\Delta/T)/(T+\theta_{NMR})$. Although the value $\Delta \sim J/10\sim 4$~K is emerging, further details of the variation of the susceptibility are not accounted for, \emph{e.g.} the relative sharpness of the variation of $\chi_{kag}$ around $T_{max}$~\cite{Lhuillier01}.

Rather than a sign of a gap, it was suggested in \cite{Mendels00} that the maximum of $\chi_{kag}$ would be the landmark of a moderate increase of magnetic correlations, later confirmed by neutron measurements~\cite{Mondelli99B}. Susceptibility calculations were further performed in kagom\'e and pyrochlore lattices with Heisenberg spins, using the so-called ``generalized constant coupling'' method (GCC)~\cite{Garcia01}. The fits for the $S=\frac{3}{2}$ kagom\'e and pyrochlore lattices, in between which the kagom\'e bilayer susceptibility is expected to lie, yield very close values of the CW temperature $\theta_{CW}=238\pm3$~K (Fig.~\ref{shift}). One extracts $J\approx37$~K for BSZCGO(0.97), consistent with the high-$T$ series expansion.

As expected, the GCC computation fits quite well the data around $T_{max}$. It only shows that the strongest underlying assumption of this cluster mean-field approach is relevant, \emph{i.e.} the spin-spin correlation length is of the order of the lattice parameter. This is also proven by neutron diffraction~\cite{Broholm90}.

The \emph{linewidth} is the second important piece of information which we extract from our data. It is indeed due to a distribution of susceptibility which is related to the existence of defects and to their nature~\cite{Limot02}. In Fig.~\ref{larg}, we compare the linewidth $\Delta H/H_{l}$ obtained for BSZCGO(0.97) to the one obtained for a comparable dilution in SCGO(0.95). In both samples, $\Delta H/H_{l}$ follows a Curie-like behavior which was shown to be characteristic of magnetic defects in SCGO$(p)$. Surprisingly, in BSZCGO(0.97), the linewidth is \emph{4 times larger} than in SCGO(0.95). 
This ratio cannot be explained by a variation of hyperfine constant, which, from the ratios of $C_{NMR}$, was estimated to a maximum of 20\%. In addition, the Ga(2d) linewidth in BSZCGO$(p)$ could be measured for various $p$ at $T=50$~K (Fig.~\ref{larg}(inset)). Contrary to SCGO$(p)$, where $\Delta H/H_{l}$ extrapolates to 0 for a perfect lattice ($p=1$), we find that it reaches an asymptotic non-zero value in BSZCGO$(p)$ for $0.86\leq p<1 $. One could wonder whether the dilution of the magnetic lattice is larger than the nominal one. 
This can be ruled out since $(i)$~we found the expected evolution of the lineshape at 300~K with $p$~\cite{BonoHFM} $(ii)$~muon spin relaxation measurements indicate a regular evolution of the dynamical properties with $p$~\cite{BonoPrep}. We then conclude that the extra width is not related to spin vacancies and reflects new \emph{$p$-independent} defects in BSZCGO$(p)$.

\begin{figure}[tb]
\center
\includegraphics[width=0.9\linewidth]{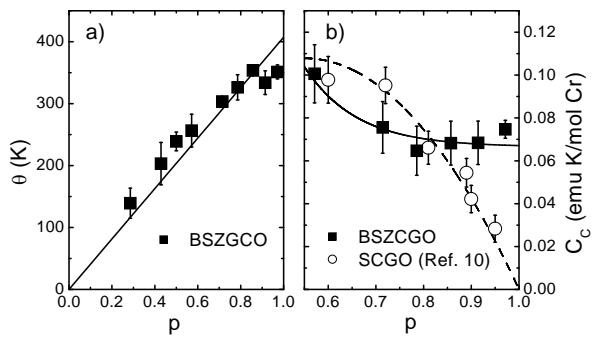}
\caption{Fits of the SQUID data (Eq.~\ref{fitSQUID}). a)~Linear fit of $\theta(p)$ for BSZCGO$(p)$. b)~$C_{C}(p)$, lines are guide to the eye. }
\label{SQUIDfits}%
\end{figure}

Defect terms could also be probed using low-$T$ macroscopic susceptibility measurements. We performed magnetization measurements under a field of 100~G, for $0.29\leq p \leq0.97$, down to 1.8~K. No difference was observed between Field Cooled and Zero Field Cooled branches down to this temperature, in agreement with \cite{Hagemann01}. Along the common route, one can simply analyze the data using the phenomenological fit~:
\begin{align}
\label{fitSQUID}
\chi_{macro}(p)=\frac{C_{CW}(p)}{T+\theta(p)} + \frac{C_{C}(p)}{T+\theta
_{C}(p)} \ .
\end{align}
The first term reflects the high-$T$ weak variation of the susceptibility due to AF interactions. From this fit, we can extract a $p$-independent effective moment per Cr$^{3+}$, $p_{eff}=4.1\pm0.2/\mu_{B}$, close to 3.87$/\mu_{B}$ expected for $S=\frac{3}{2}$. From the linear variation of $\theta(p) \approx 1.5\,\theta_{CW}(p)$ (Fig.~\ref{SQUIDfits}(a)), we extract $J\sim40$~K, consistent with our NMR result.

The second term in Eq.~\ref{fitSQUID} is introduced to fit the low-$T$ behavior associated with the defects~\cite{Schiffer97,Limot02}. It is close to pure Curie law, with $\theta_{C}(p)=1\pm0.5$~K. Values of the Curie constants $C_{C}$ are reported in Fig.~\ref{SQUIDfits}(b). We find a \emph{$p$-independence} qualitatively similar to the NMR width data and a finite value for $p\rightarrow 1$. To give an order of magnitude, this term represents 20\% of $S=\frac{1}{2}$ paramagnetic spins with respect to the number of Cr$^{3+}$ spins. Once the dilution defect term becomes significant ($p\leq0.8$), the $C_{C}(p)$ slope is qualitatively comparable in both systems~\footnote{The crossing between the variation of $C_{C}(p)$ of both systems (Fig.~\ref{SQUIDfits}(b)) is not observed in the NMR width (Fig.~\ref{larg}(inset)). We attribute this difference to the extra Ga-Cr substitutions in the Cr(4f$_{vi}$) sites in SCGO, measured in $\chi_{macro}$ and not in $K_{2d}^{71}\propto\chi_{kag}$ (\emph{unpublished}).}.

To summarize, low-$T$ $\chi_{macro}$ and NMR width data can be both satisfactorily accounted for, only if a novel $p$-independent defect-like contribution is taken into consideration, a result somehow
surprising in view of the close similarity between kagom\'e bilayers. We now elaborate about the nature of these defects. The only major change lies in the 1:1 random occupancy of the 2d site by Ga$^{3+}$ or Zn$^{2+}$ ions which induces different electrostatic interactions with the neighboring ions. As an example, distances to O$^{2-}$ in a tetrahedral environment vary from $r_{Ga^{3+}-O^{2-}}=0.47$~\AA\ to $r_{Zn^{2+}-O^{2-}}=0.60$~\AA\ which is at the origin of a change of the average Ga(4f,2d)-O bonds, from 1.871~\AA\ for SCGO~\cite{Lee96} to 1.925~\AA\ in BSZCGO~\cite{Hagemann01}. Similarly, one expects that the Cr$^{3+}$ will be less repelled by Zn$^{2+}$ than by Ga$^{3+}$, which certainly induces magnetic bond-disorder, \emph{i.e.} a modulation of exchange interactions $J$ between neighboring Cr$^{3+}$. As seen before, one can estimate that a modulation as low as $0.01$~\AA\ in the Cr$^{3+}$-Cr$^{3+}$ distances would yield a $10$\% modulation of $J$.

The presence of non-perfect equilateral triangles classically induces a paramagnetic component in the susceptibility of the frustrated units~\cite{Moessner99}. The fact that we do not observe any extra Curie variation as compared to SCGO$(p)$ in the average susceptibility, probed through the NMR shift $K_{2d}^{71}$ (Fig.~\ref{shift}), indicates that such unbalanced exchange interactions only induce a staggered response in the same manner as spin vacancies~\cite{Limot02}. Whether this could be connected with localization of singlets in the vicinity of defects~\cite{Dommange03} and a staggered cloud around should be more deeply explored. This might highlight the relevance of a RVB approach to the physics of the kagom\'e network. Further insight into the exact nature of the defects likely requires a better determination of the bond disorder through structural studies at low-$T$, to avoid the usual thermal fluctuations at room $T$.

To conclude, one of our major findings is the maximum of the local susceptibility $\chi_{kag}$ observed around 45~K for \emph{both} BSZCGO$(p)$ and SCGO$(p)$, which appears very robust to the presence of defects either
spin vacancies as observed in \cite{Limot02} or bond disorder as pointed out above. This intrinsic susceptibility is very close to a model of $S=\frac{3}{2}$ kagom\'e or pyrochlore with correlation length of the order of magnitude of the Cr$^{3+}$-Cr$^{3+}$ bond. Finally, we showed that bond disorder is likely to give raise, like spin vacancies, to extended defects which underline the correlated nature of the kagom\'e network. This study of ``impurities'' might be a major avenue for revealing the nature of the ground state in a similar manner as in low-D quantum magnetism studies.

We thank F.~Bert, J.~Bobroff, A.~Keren, C.~Lhuillier, R.~Moessner, C.~Payen and P.~Sindzingre for discussions.


\begin{thebibliography}{0}
\expandafter\ifx\csname natexlab\endcsname\relax\def\natexlab#1{#1}\fi
\expandafter\ifx\csname bibnamefont\endcsname\relax
  \def\bibnamefont#1{#1}\fi
\expandafter\ifx\csname bibfnamefont\endcsname\relax
  \def\bibfnamefont#1{#1}\fi
\expandafter\ifx\csname citenamefont\endcsname\relax
  \def\citenamefont#1{#1}\fi
\expandafter\ifx\csname url\endcsname\relax
  \def\url#1{\texttt{#1}}\fi
\expandafter\ifx\csname urlprefix\endcsname\relax\def\urlprefix{URL }\fi
\providecommand{\bibinfo}[2]{#2}
\providecommand{\eprint}[2][]{\url{#2}}

\end{thebibliography}


\begin{thebibliography}{99}
\expandafter\ifx\csname natexlab\endcsname\relax\def\natexlab#1{#1}\fi
\expandafter\ifx\csname bibnamefont\endcsname\relax
  \def\bibnamefont#1{#1}\fi
\expandafter\ifx\csname bibfnamefont\endcsname\relax
  \def\bibfnamefont#1{#1}\fi
\expandafter\ifx\csname citenamefont\endcsname\relax
  \def\citenamefont#1{#1}\fi
\expandafter\ifx\csname url\endcsname\relax
  \def\url#1{\texttt{#1}}\fi
\expandafter\ifx\csname urlprefix\endcsname\relax\def\urlprefix{URL }\fi
\providecommand{\bibinfo}[2]{#2}
\providecommand{\eprint}[2][]{\url{#2}}

\bibitem[{\citenamefont{Ramirez}(2001)}]{Ramirez01} A.P.~Ramirez, in \emph{Handbook on Magnetic Materials}, edited by K.J.H.~Busch, \textbf{13}, 423 (Elsevier Science, Amsterdam, 2001).

\bibitem[{\citenamefont{Mambrini et~al.}(2000)}]{Mambrini00}
\bibinfo{author}{\bibfnamefont{M.}~\bibnamefont{Mambrini}},
  \bibinfo{author}{\bibfnamefont{F.}~\bibnamefont{Mila}},
  \bibinfo{journal}{{E}ur.\ {P}hys.\ {J}.\ {B}} \textbf{\bibinfo{volume}{17}},
  \bibinfo{pages}{651} (\bibinfo{year}{2000}).

  \bibitem[{\citenamefont{Diep}(2003)}]{Diep03} \emph{Frustrated Spin Systems}, edited by H.T.~Diep (World Scientific, 2003).

\bibitem[{\citenamefont{Anderson}(2001)}]{Anderson87}
\bibinfo{author}{\bibfnamefont{P.W.}~\bibnamefont{Anderson}},
  \bibinfo{journal}{{S}cience} \textbf{\bibinfo{volume}{235}},
  \bibinfo{pages}{1196} (\bibinfo{year}{1987}).

\bibitem[{\citenamefont{Baskaran}(2003)}]{Baskaran03}
\bibinfo{author}{\bibfnamefont{G.}~\bibnamefont{Baskaran}} \emph{et al.},
  \bibinfo{journal}{{P}hys.\ {R}ev.\ {L}ett.} \textbf{\bibinfo{volume}{91}},
  \bibinfo{pages}{097003} (\bibinfo{year}{2003}).

\bibitem[{\citenamefont{Chubukov}(1992)}]{Chubukov92}
\bibinfo{author}{\bibfnamefont{A.}~\bibnamefont{Chubukov}},
  \bibinfo{journal}{{P}hys.\ {R}ev.\ {L}ett.} \textbf{\bibinfo{volume}{69}},
  \bibinfo{pages}{832} (\bibinfo{year}{1992});
\bibinfo{author}{\bibfnamefont{R.}~\bibnamefont{Moessner}},
  \bibinfo{journal}{{P}hys.\ {R}ev.\ {B}} \textbf{\bibinfo{volume}{57}},
  \bibinfo{pages}{R5587} (\bibinfo{year}{1998});
\bibinfo{author}{\bibfnamefont{S.E.}~\bibnamefont{Palmer}} \bibnamefont{and}
  \bibinfo{author}{\bibfnamefont{J.T.}~\bibnamefont{Chalker}},
  \bibinfo{journal}{{P}hys.\ {R}ev.\ {B}} \textbf{\bibinfo{volume}{62}},
  \bibinfo{pages}{488} (\bibinfo{year}{2000});
\bibinfo{author}{\bibfnamefont{M.}~\bibnamefont{Elhajal}},
  \bibinfo{author}{\bibfnamefont{B.}~\bibnamefont{Canals}}, \bibnamefont{and}
  \bibinfo{author}{\bibfnamefont{C.}~\bibnamefont{Lacroix}},
  \bibinfo{journal}{{P}hys.\ {R}ev.\ {B}} \textbf{\bibinfo{volume}{66}},
  \bibinfo{pages}{014422} (\bibinfo{year}{2002}).

\bibitem[{\citenamefont{Hagemann et~al.}(2001)}]{Hagemann01}
\bibinfo{author}{\bibfnamefont{I.S.}~\bibnamefont{Hagemann}} \emph{et al.},
  \bibinfo{journal}{{P}hys.\ {R}ev.\ {L}ett.} \textbf{\bibinfo{volume}{86}},
  \bibinfo{pages}{894} (\bibinfo{year}{2001}).

\bibitem[{\citenamefont{Bonnet et~al.}(2004)}]{BonnetHFM}
\bibinfo{author}{\bibfnamefont{P.}~\bibnamefont{Bonnet}} \emph{et al.},
  \bibinfo{journal}{{J}. {P}hys.: {C}ondens. {M}atter } \textbf{\bibinfo{volume}{16}},
  \bibinfo{pages}{S835} (\bibinfo{year}{2004}).

\bibitem[{\citenamefont{Bono et~al.}(2003)}]{BonoPrep} D.~Bono \emph{et al.}, in preparation.

\bibitem[{\citenamefont{Limot et~al.}(2002)}]{Limot02}
\bibinfo{author}{\bibfnamefont{L.}~\bibnamefont{Limot}} \emph{et al.},
  \bibinfo{journal}{{P}hys.\ {R}ev.\ {B}} \textbf{\bibinfo{volume}{65}},
  \bibinfo{pages}{144447} (\bibinfo{year}{2002}).

\bibitem[{\citenamefont{Obradors et~al.}(1988)}]{Obradors88}
\bibinfo{author}{\bibfnamefont{X.}~\bibnamefont{Obradors}} \emph{et al.},
  \bibinfo{journal}{{S}olid\ {S}tate\ {C}ommun.} \textbf{\bibinfo{volume}{65}},
  \bibinfo{pages}{189} (\bibinfo{year}{1988}).

\bibitem[{\citenamefont{Lee et~al.}(1996)}]{Lee96}
\bibinfo{author}{\bibfnamefont{S.-H.} \bibnamefont{Lee}} \emph{et al.},
  \bibinfo{journal}{{P}hys.\ {R}ev.\ {L}ett.} \textbf{\bibinfo{volume}{76}},
  \bibinfo{pages}{4424} (\bibinfo{year}{1996}).

\bibitem[{\citenamefont{Mendels et~al.}(2000)}]{Mendels00}
\bibinfo{author}{\bibfnamefont{P.}~\bibnamefont{Mendels}} \emph{et al.},
  \bibinfo{journal}{{P}hys.\ {R}ev.\ {L}ett.} \textbf{\bibinfo{volume}{85}},
  \bibinfo{pages}{3496} (\bibinfo{year}{2000}).

\bibitem[{\citenamefont{Keren et~al.}(1998)\citenamefont{Keren, Mendels,
  Horvati\'c, Ferrer, Uemura, Mekata, and Asano}}]{Keren98}
\bibinfo{author}{\bibfnamefont{A.}~\bibnamefont{Keren}} \emph{et al.},
  \bibinfo{journal}{{P}hys.\ {R}ev.\ {B}} \textbf{\bibinfo{volume}{57}},
  \bibinfo{pages}{10745} (\bibinfo{year}{1998}).

\bibitem[{\citenamefont{Harris et~al.}(1992)\citenamefont{Harris, Kallin, and
  Berlinsky}}]{Harris92}
\bibinfo{author}{\bibfnamefont{A.B.}~\bibnamefont{Harris}},
  \bibinfo{author}{\bibfnamefont{C.}~\bibnamefont{Kallin}}, \bibnamefont{and}
  \bibinfo{author}{\bibfnamefont{A.J.}~\bibnamefont{Berlinsky}},
  \bibinfo{journal}{{P}hys.\ {R}ev.\ {B}} \textbf{\bibinfo{volume}{45}},
  \bibinfo{pages}{2899} (\bibinfo{year}{1992}).

\bibitem[{\citenamefont{Waldtmann et~al.}(1999)}]{Waldtmann98}
\bibinfo{author}{\bibfnamefont{Ch.}~\bibnamefont{Waldtmann}} \emph{et al.},
  \bibinfo{journal}{{E}ur.\ {P}hys.\ {J}.\ {B}} \textbf{\bibinfo{volume}{2}},
  \bibinfo{pages}{501} (\bibinfo{year}{1998}).

\bibitem[{\citenamefont{Lhuillier}(2001)}]{Lhuillier01} C.~Lhuillier and P.~Sindzingre, in \emph{Quantum Properties of Low-Dimensional Antiferromagnets}, edited by Y.~Ajiro and J.-P.~Boucher (Kyushu University Press, 2001).

\bibitem[{\citenamefont{Mondelli et~al.}(1999)}]{Mondelli99B}
\bibinfo{author}{\bibfnamefont{C.}~\bibnamefont{Mondelli}} \emph{et al.},
  \bibinfo{journal}{{P}hysica\ {B}} \textbf{\bibinfo{volume}{267-268}},
  \bibinfo{pages}{139} (\bibinfo{year}{1999}).

\bibitem[{\citenamefont{Garcia-Adeva and Huber}(2001)}]{Garcia01}
\bibinfo{author}{\bibfnamefont{A.J.}~\bibnamefont{Garc\'\i a-Adeva}}
  \bibnamefont{and} \bibinfo{author}{\bibfnamefont{D.L.}~\bibnamefont{Huber}},
  \bibinfo{journal}{{P}hys.\ {R}ev.\ {B}} \textbf{\bibinfo{volume}{63}},
  \bibinfo{pages}{174433} (\bibinfo{year}{2001});
  \bibinfo{journal}{{P}hysica {B}} \textbf{\bibinfo{volume}{320}},
  \bibinfo{pages}{18} (\bibinfo{year}{2002}).

\bibitem[{\citenamefont{Broholm et~al.}(1990)\citenamefont{Broholm, Aeppli,
  Espinosa, and Cooper}}]{Broholm90}
\bibinfo{author}{\bibfnamefont{C.}~\bibnamefont{Broholm}},
  \bibinfo{author}{\bibfnamefont{G.}~\bibnamefont{Aeppli}},
  \bibinfo{author}{\bibfnamefont{G.P.}~\bibnamefont{Espinosa}}, \bibnamefont{and}
  \bibinfo{author}{\bibfnamefont{A.S.}~\bibnamefont{Cooper}},
  \bibinfo{journal}{{P}hys.\ {R}ev.\ {L}ett.} \textbf{\bibinfo{volume}{65}},
  \bibinfo{pages}{3173} (\bibinfo{year}{1990}).

\bibitem[{\citenamefont{Bono et~al.}(2004)}]{BonoHFM}
\bibinfo{author}{\bibfnamefont{D.}~\bibnamefont{Bono}} \emph{et al.},
  \bibinfo{journal}{{J}. {P}hys.: {C}ondens. {M}atter } \textbf{\bibinfo{volume}{16}},
  \bibinfo{pages}{S817} (\bibinfo{year}{2004}).

\bibitem[{\citenamefont{Schiffer and Daruka}(1997)}]{Schiffer97}
\bibinfo{author}{\bibfnamefont{P.}~\bibnamefont{Schiffer}},
  \bibnamefont{and} \bibinfo{author}{\bibfnamefont{I.}~\bibnamefont{Daruka}},
  \bibinfo{journal}{{P}hys.\ {R}ev.\ {B}} \textbf{\bibinfo{volume}{56}},
  \bibinfo{pages}{13712} (\bibinfo{year}{1997}).
  
\bibitem[{\citenamefont{Moessner and Berlinsky}(1999)}]{Moessner99}
\bibinfo{author}{\bibfnamefont{R.}~\bibnamefont{Moessner}} \bibnamefont{and}
  \bibinfo{author}{\bibfnamefont{A.J.}~\bibnamefont{Berlinsky}},
  \bibinfo{journal}{{P}hys.\ {R}ev.\ {L}ett.} \textbf{\bibinfo{volume}{83}},
  \bibinfo{pages}{3293} (\bibinfo{year}{1999}).

\bibitem[{\citenamefont{Dommange et al.}(2003)}]{Dommange03}
\bibinfo{author}{\bibfnamefont{S.}~\bibnamefont{Dommange}},
  \bibinfo{author}{\bibfnamefont{M.}~\bibnamefont{Mambrini}},
  \bibinfo{author}{\bibfnamefont{B.}~\bibnamefont{Normand}}, \bibnamefont{and}
  \bibinfo{author}{\bibfnamefont{F.}~\bibnamefont{Mila}}, 
  \bibinfo{journal}{{P}hys.\ {R}ev.\ {B}} \textbf{\bibinfo{volume}{68}},
  \bibinfo{pages}{224416} (\bibinfo{year}{2003}).

\end{thebibliography}
\end{document}